\documentclass[twocolumn,shortnote]{jpsj2} 

\def\runauthor{H. Fukazawa {\it et al.}}

\title{%
Suppression of Magnetic Order\\
by Pressure in BaFe$_{2}$As$_{2}$ 
}

\author{%
Hideto~Fukazawa\thanks{E-mail address: hideto@nmr.s.chiba-u.ac.jp}, 
Nao~Takeshita$^{1}$,
Takehiro~Yamazaki,
Kenji~Kondo,
Kenji~Hirayama,
Yoh~Kohori, 
Kiichi~Miyazawa$^{1,2}$,
Hijiri~Kito$^{1}$, 
Hiroshi~Eisaki$^{1}$, 
and Akira~Iyo$^{1}$
}

\inst{%
Department of Physics, Chiba University, Chiba 263-8522\\
$^{1}$Nanoelectronics Research Institute, National Institute of 
Advanced Industrial Science and Technology, Tsukuba 305-8562\\
$^{2}$Department of Applied Electronics, Tokyo University of Science, Noda 278-8510
}

\recdate{\today}

\kword{%
BaFe$_{2}$As$_{2}$, magnetic transition, superconductivity, high pressure, nuclear magnetic resonance
}

\begin{document}
\maketitle


The discovery of superconductivity in F-doped LaFeAsO~\cite{Kam1} 
has accelerated further investigations of the related superconductors~\cite{Kit1,Take1,Ren2}. 
The common feature of these compounds is the possession of a FeAs layer which is analogous to the CuO$_{2}$ plane 
in high-superconducting-transition-temperature (high-$T_{\rm c}$) cuprates. 
Soon after the intensive investigations of the oxypnictides, 
the oxygen-free iron pnictide BaFe$_{2}$As$_{2}$~\cite{Rot1} 
was proposed as the next candidate parent material of superconductor with high $T_{\rm c}$. 
The crystal structure of this pnictide is the ThCr$_{2}$Si$_{2}$-type structure and 
possesses a similar FeAs layer to that realized in LaFeAsO. 
Moreover, this material exhibits spin density wave (SDW) anomaly at $T_{\rm SDW} =$~140~K (BaFe$_{2}$As$_{2}$)~\cite{Rot1,Hua1}. 
It is important to note that the compound exhibits structural phase transition 
from tetragonal ($I4/mmm$) to orthorhombic ($Fmmm$) as well as SDW anomaly~\cite{Rot1,Hua1}.  
The most striking feature of this compound is that the SDW anomaly disappears and the superconductivity indeed sets in 
by hole doping, for example, K substitution for Ba~\cite{Rot2}. 
In addition, the pressure $P$ induced superconductivity in BaFe$_{2}$As$_{2}$ was recently 
reported by the magnetic susceptibility measurements under pressure up to about 6~GPa~\cite{Ali1}. 
Although it should be revealed by additional experimental approaches whether this superconductivity is intrinsic or not, 
it is clear that the investigation of this compound under pressure is quite important. 
In order to understand the relation between the SDW instability and expected superconductivity in BaFe$_{2}$As$_{2}$, 
we performed resistivity $\rho$ measurements and zero-external-field (ZF) 
$^{75}$As nuclear-magnetic-resonance (NMR) measurements of BaFe$_{2}$As$_{2}$ under high pressure. 

In our previous letter, we reported the $^{75}$As-NMR spectra of BaFe$_{2}$As$_{2}$ at various temperatures 
and the temperature $T$ dependence of the spin-lattice relaxation rate $1/T_{1}$ of $^{75}$As under ambient pressure~\cite{Fuk2}. 
$^{75}$As-NMR spectra clearly revealed that magnetic transition occurs at around 131~K in our samples, 
which corresponds to the emergence of spin density wave. 
The $T$ dependence of the internal magnetic field $H_{\rm int}$ suggested that the transition is likely of the first order. 
We also deduced from the difference of the linewidth of the highly oriented powder and partially oriented powder 
that the $H_{\rm int}$ at the As site is parallel to $c$-axis. 
The steep decrease in $1/T_{1}$ below $T_{\rm SDW}$ was observed and was due to the gap formation of the SDW at part of the Fermi surface. 
$1/T_{1}$ below about 100~K was nearly proportional to $T$ which indicates 
the relaxation due to the remaining conduction electron at the Fermi level even below $T_{\rm SDW}$. 
In our previous paper, we also reported a ZF $^{75}$As-NMR spectrum at 1.5~K~\cite{Fuk2,Fuk3}. 
The ZF $^{75}$As-NMR spectrum clearly revealed that the magnetically ordered state of this compound is commensurate. 
Furthermore, the center line of the ZF $^{75}$As-NMR spectrum is a good measure of the $H_{\rm int}$ at the As site.


The polycrystalline BaFe$_{2}$As$_{2}$ was synthesized by a high-temperature and high-pressure method. 
The samples were confirmed to be nearly single phase by X-ray diffraction analysis. 
The electrical resistivity was measured by means of ordinary dc four probe method. 
Cubic anvil apparatus for low temperature measurement~\cite{Mor1} was used for the resistivity measurements under $P$ up to 13~GPa. 
Detailed condition for the use of this apparatus is given elsewhere~\cite{Take1}. 
The NMR experiment on the $^{75}$As nucleus ($I=3/2$, $\gamma = 7.292$~MHz/T) was carried out 
using a phase-coherent pulsed NMR spectrometer. 
The samples were crushed into powder for use in the experiment. 
The CuBe-NiCrAl hybrid piston-cylinder cell was used for the NMR measurements under $P$ up to about 2.5~GPa. 
Daphne 7373 was used for the pressure transmitting medium, which was confirmed to be 
suited for the measurements under $P$ with the use of the piston cylinder cell~\cite{Fuk4}. 
The pressure and its homogeneity was determined 
from the nuclear quadrupole resonance frequency and full width at half maximum of $^{63}$Cu of Cu$_{2}$O at 4.2~K, respectively~\cite{Rey1}.


 \begin{figure}
  \centering
  \includegraphics[width=8cm]{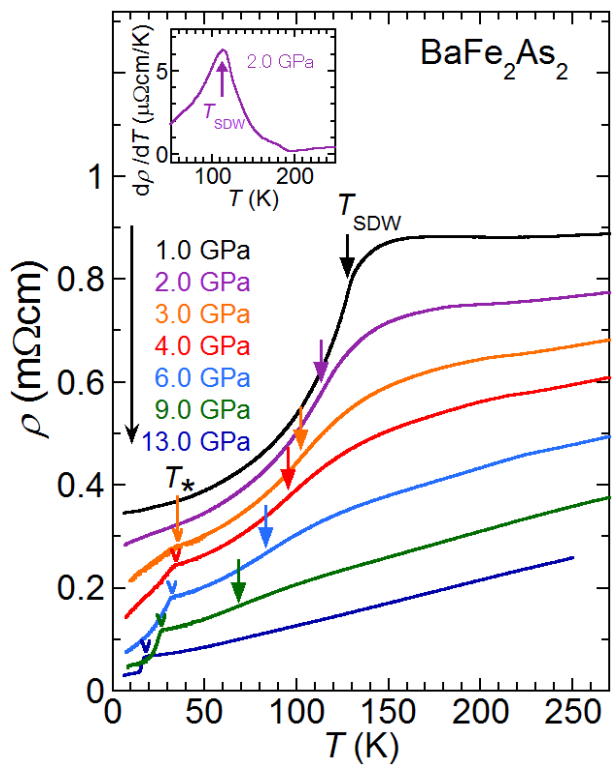}
  \caption{
  (Color online) Temperature $T$ dependence of the dc resistivity $\rho$ of BaFe$_{2}$As$_{2}$ under various pressures. 
  The inset shows the $T$ dependence of $d\rho /dT$ of BaFe$_{2}$As$_{2}$ at 2.0~GPa.
  }
 \end{figure}

In Fig.~1, we show the $T$ dependence of $\rho$ of BaFe$_{2}$As$_{2}$ under various pressures. 
$\rho$ exhibits metallic behavior in almost all the $T$ range at all the pressures. 
At ambient pressure, $\rho$ rapidly decreases below about 131~K (= $T_{\rm SDW}$)~\cite{Tom1}. 
This is due to SDW anomaly as well as the structural phase transition. 
The $T_{\rm SDW}$ of our samples is slightly lower than those 
reported by Rotter {\it et al.}~\cite{Rot1} and Huang {\it et al.}~\cite{Hua1} 
This anomaly apparently becomes broader above 1~GPa: it becomes difficult to determine the SDW and/or structural transition temperature. 
In the inset of Fig.~1, we show the $T$ dependence of $d\rho /dT$ of BaFe$_{2}$As$_{2}$ at 2.0~GPa.
The large peak can be seen at around 114~K. 
Because the $\rho$ decreases more steeply below this temperature, 
which can be interpreted as the suppression of the magnetic scattering, 
this anomaly is attributable to SDW transition. 
Hence, we express the characteristic temperature of this anomaly under pressure with $T_{\rm SDW}$ as well as the ambient $T_{\rm SDW}$. 
Gradual decrease of $\rho$ at the temperatures above $T_{\rm SDW}$ might be due to the preceding behavior of SDW anomaly. 

 \begin{figure}
  \centering
  \includegraphics[width=8cm]{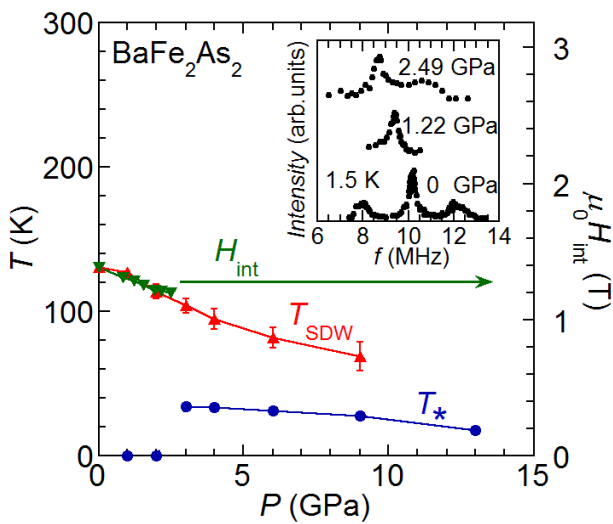}
  \caption{
  (Color online) The $T$-$P$ phase diagram of BaFe$_{2}$As$_{2}$ determined from resistivity anomalies. 
  Squares ($T_{\rm anom}$), up triangles ($T_{\rm SDW}$), and circles ($T_{\rm *}$)
  denote structural-like, SDW, and SC-like transition, respectively. 
  Down triangles denote the internal magnetic fields $H_{\rm int}$ at the As site evaluated from the ZF $^{75}$As-NMR measurements at 1.5~K. 
  The inset shows the ZF $^{75}$As-NMR spectra at various pressures. 
  }
 \end{figure}

More clear change in $\rho$ is steep decrease below about 20-30~K under $P$ above 3.0~GPa. 
This anomaly corresponds to superconducting phase transition 
which was reported by the magnetic susceptibility measurements under $P$ up to about 6~GPa~\cite{Ali1}. 
However, we should note that we obtained no zero-resistivity at every pressure. 
We confirmed its reproducibility with different batches of samples. 
Simultaneously, we obtained current dependence of the $\rho$ below these temperatures. 
Therefore, this anomaly is related with superconductivity, 
but we could not conclude that this anomaly is an intrinsic property of BaFe$_{2}$As$_{2}$ under pressure. 
Then, we express the characteristic temperature of this anomaly with $T_{\rm *}$.


In Fig.~2, we show the $T$-$P$ phase diagram of BaFe$_{2}$As$_{2}$ determined from resistivity anomalies. 
We also show the $P$ dependence of the $H_{\rm int}$ up to about 2.5~GPa. 
The $H_{\rm int}$ at the As site was evaluated from the $P$-dependent center line 
of the ZF $^{75}$As-NMR spectrum as shown in the inset of Fig.~2. 
Detailed analysis of the spectra to evaluate the $H_{\rm int}$ is written elsewhere~\cite{Fuk3}. 
We also show higher satellite line at 2.49~GPa in the inset of Fig.~2 
in order to show that the peak at around 8.7~MHz is indeed a center line of the ZF NMR spectrum. 
The $P$ dependence of the $H_{\rm int}$ is nearly proportional to that of $T_{\rm SDW}$. 
This supports that our estimation of the $T_{\rm SDW}$ from $\rho$ is adequate. 
The phase diagram clearly shows that the SDW anomaly is suppressed by applied $P$, but is simultaneously quite robust against $P$.  
Furthermore, if $T_{*}$ is an intrinsic $T_{c}$ of BaFe$_{2}$As$_{2}$, 
this phase diagram suggests the possibility of coexistence of SDW ordered phase and superconducting phase. 
The similar behavior is reported by the phase diagram of the isomorph SrFe$_{2}$As$_{2}$ under $P$~\cite{Kum1} 
and the K-doped BaFe$_{2}$As$_{2}$~\cite{HCh1}. 
However, we again emphasize that, at the present stage, we cannot conclude that the anomaly below about 20-30~K is 
an intrinsic property of BaFe$_{2}$As$_{2}$ under $P$. 
We note that there is no clear change in spin-lattice relaxation time $T_{1}$ at 1.5~K between 0 and 2.49~GPa. 
If superconductivity sets in even at 2.49~GPa, we can expect the sufficient increase of $T_{1}$ due to the gap formation of superconductivity. 
This might indicate that the apparent superconductivity above 3~GPa is not a bulk superconductivity. 
The imperfection of the crystal possibly suppresses the superconductivity or 
the anomaly arises from a small amount of impurity phase consisting of Fe and As, which cannot be detected by NMR measurement. 
Therefore, we should continue the NMR measurements under $P$ beyond 3~GPa 
in order to determine the magnetic phase diagram and ground state of BaFe$_{2}$As$_{2}$ under $P$ 
with Bridgman anvil cell~\cite{Fuk1} and mini cubic anvil apparatus~\cite{Hir1}, which are now in progress.




This work is supported by a Grant-in-Aid for Scientific Research from the MEXT.

\end{document}